\documentclass[paper]{aa}
\usepackage{graphicx}
\usepackage{txfonts}
\def\swift{{\em Swift\/~}}
\def\einstein{{\em Einstein\/~}}
\def\sax{{\em Beppo}SAX\/~}

\def\ergcms{\mbox{ erg cm$^{-2}$ s$^{-1}$}}
\def\pks{PKS 0548$-$322}

\begin{document}
\title{Swift XRT and UVOT deep observations of the high energy peaked \\ BL Lac object PKS~0548$-$322 
       close to its brightest state}
\author{M.~Perri\inst{1}, A.~Maselli\inst{2}, P.~Giommi\inst{1,3}, 
        E.~Massaro\inst{2}, R.~Nesci\inst{2}, A.~Tramacere\inst{2}, M.~Capalbi\inst{1}, 
        G.~Cusumano\inst{4}, G.~Chincarini\inst{5,6}, G.~Tagliaferri\inst{5}, 
        D.N.~Burrows\inst{7}, D.V.~Berk\inst{7}, N.~Gehrels\inst{8}, R.M.~Sambruna\inst{8}
       }
   \institute{ASI Science Data Center, Via Galileo Galilei, 
              I-00044 Frascati, Italy
	\and Dipartimento di Fisica, Universit\`a La Sapienza, Piazzale A. Moro 2, 
	     I-00185 Roma, Italy
	\and Agenzia Spaziale Italiana, Unit\`a Osservazione dell'Universo,
             Viale Liegi 26, I-00198 Roma, Italy
	 \and INAF -- Istituto di Astrofisica Spaziale e Fisica Cosmica,
              Sezione di Palermo, Via La Malfa 153, I-90146 Palermo, Italy
	 \and INAF -- Osservatorio Astronomico di Brera, 
              Via Bianchi 46, I-23807 Merate, Italy
	 \and Universit\`a degli Studi di Milano-Bicocca, Dipartimento 
              di Fisica, Piazza delle Scienze 3, I-20126 Milano, Italy
	 \and Department of Astronomy \& Astrophysics, Pennsylvania State 
              University, University Park, PA 16802, USA
	 \and NASA/Goddard Space Flight Center, Greenbelt, MD 20771, USA
}

\offprints{\email{perri@asdc.asi.it}}
\date{Received / Accepted}

\titlerunning{\swift XRT and UVOT deep observations of the BL Lac object PKS~0548$-$322}
\authorrunning{M. Perri et al.}

\abstract
%context heading (optional), leave it empty if necessary
{}
%aims heading (mandatory)
{We observed the high energy peaked BL Lac object \pks~ (BZB J0550$-$3216) with \swift to study the 
temporal and spectral properties of its synchrotron emission simultaneously in the Optical, Ultraviolet and 
X-ray energy bands.  
}
%methods  heading (mandatory)
{We carried out a spectral analysis of 5 \swift XRT and UVOT observations 
of \pks~ over the period April - June 2005.
}
%results heading (mandatory)
{The X-ray flux of this BL Lac source was found to be approximately 
constant at a level of $F_{(2-10~{\mathrm keV})}\simeq 4\times 10^{-11}$ \ergcms, a factor 
of 2 brighter than when observed by \sax in 1999 and close to the maximum intensity 
reported in the \einstein Slew Survey.
The very good statistics obtained in the 0.3-10 keV \swift X-ray spectrum allowed us to detect 
highly significant deviations from a simple power law spectral distribution. A log-parabolic 
model describes well the X-ray data and gives a best fit curvature parameter of 0.18 and a peak 
energy in the Spectral Energy Distribution of about 2 keV. The UV spectral data from \swift UVOT 
join well with a power law extrapolation of the soft X-ray data points suggesting that the same component 
is responsible for the observed emission in the two bands. The combination 
of synchrotron peak in the X-ray band and high intensity state confirms \pks~ as a prime target 
for TeV observations. X-ray monitoring and coordinated TeV campaigns are highly advisable.
}
%conclusion heading (optional), leave it empty if necessary 
{}

\maketitle

\section{Introduction}
\label{intro}

BL Lacertae objects (BL Lacs) constitute a  rather peculiar and extreme class of Active Galactic Nuclei
(AGNs). The main characteristics that distinguish BL Lacs are the rapid variability at all 
frequencies, high and variable radio and optical polarization, compact and flat-spectrum radio 
emission, superluminal motion, smooth and broad non-thermal continuum covering the electromagnetic 
spectrum from radio to $\gamma$-rays and the almost complete absence of emission lines in the optical 
band.
The extreme properties of BL Lacs are successfully explained in terms of relativistic beaming, 
i.e. of a relativistic bulk motion of the emitting region toward the observer 
(\cite{Blandford78}; \cite{Urry95}).

The Spectral Energy Distribution (SED) of BL Lacs is generally characterized, 
in a \mbox{$Log(\nu\,F(\nu))$--$Log(\nu)$} 
representation, by two emission peaks: the first is produced by synchrotron emission by 
relativistic electrons in a jet closely aligned to the line of sight, while inverse Compton 
scattering by the same population of relativistic electrons is thought to be at the origin of the higher 
energy peak (e.g. \cite{Ghisellini89}).
BL Lacs are often divided into two classes according to the position of the synchrotron energy peak: 
low energy peaked BL Lacs (LBLs), with the peak located at IR/optical wavelengths, and high 
energy peaked BL Lacs (HBLs) with the synchrotron emission peaking in the UV/X-ray energy band 
(\cite{Giommi94}; \cite{Padovani95}).

\pks~ ($z$=0.069, \cite{fosdis76}), also named BZB J0550$-$3216 in the
recent Multifrequency Catalogue of Blazars (\cite{Massaro05}), is 
a remarkable BL Lac object characterized by a relatively strong and fast 
variability in the X-ray energy band (\cite{Blustin04}). 
It is hosted in a giant elliptical galaxy (Falomo et al. 1995, \cite{Wurtz96}) 
which is the dominant member of a rich cluster of galaxies.
The synchrotron power peaks in the X-ray band and for this reason
it is classified as an HBL source (\cite{Padovani95}).
\pks~ was observed on several occasions by a number of X-ray astronomical satellites
showing strong variations both in intensity and in spectral shape. 
A brief summary of historical X-ray data on \pks~ can be found in \cite{Cost01a} 
who also presented the analysis of three {\it Beppo}SAX observations. 
The brightest flux was reported in the {\it Einstein} Slew Survey (\cite{Per96b}). 
The possibility that the X-ray spectrum of \pks~ might deviate from a simple power 
law was apparent since the very early observations.
\cite{Urry86}, using {\it Einstein} data, interpreted the observed curvature as 
due to an excess low energy absorption, whereas \cite{Madejski85} suggested a spectral 
bending.  
\begin{table}[t]
\caption{\swift observations and exposures of \pks~ in spring 2005.}
\label{tab1}
\begin{center}
\begin{tabular}{lcrc}
\hline
 Date     & Start UT & XRT Exp. &  UVOT Exp. \\
          &          &      (s)~~~~~     &    (s)~~~~~  \\
\hline
April 1  & 00:43 & 1,258~~~   & /~          \\
April 26 & 23:31 & 5,212~~~   & 7,791~~~  \\
April 28 & 10:53 & 1,354~~~   & /~          \\
May 13   & 14:22 & 3,362~~~   & 3,380~~~  \\
May 21   & 10:09 & 9,224~~~  & 9,135~~~  \\
May 22   & 22:52 & 40,191~~~  & /~          \\
May 24   & 00:02 & 1,349~~~   & 5,402~~~  \\
May 26   & 00:46 &   400~~~ & 1,926~~~  \\
May 29   & 00:07 & 8,083~~~  & 8,715~~~  \\
June 14  & 16:23 & 1,235~~~  & /~          \\
June 24  & 17:15 & 7,951~~~  & /~          \\
\hline
\multicolumn{4}{c} { }
\end{tabular}
\end{center}
\end{table}
\pks~ was observed by {\it EXOSAT} at five epochs between 1983 and 1986 (\cite{Barr88}; 
\cite{Garilli90}) and broken power law best fits gave a peak energy of the Spectral Energy 
Distribution in the 2.5--5 keV range.
Similar results were obtained in the re-analysis of all {\it EXOSAT} observations by 
\cite{Ghosh95}, who noticed that a log-parabolic law represents 
well the multifrequency spectrum of \pks.
A subsequent {\it GINGA} observation in February 1991 (\cite{Tashiro95}) showed a 
much flatter spectrum, with an X-ray photon index of 1.84$\pm$0.02 in the 2--30 
keV range implying that the synchrotron peak in the SED had moved at energies larger 
than 30 keV.
\pks~ was observed twice by \sax in 1999 on February 20 and on April 7. 
A first analysis was reported by \cite{Cost01a} who found evidence of intrinsic curvature 
because a broken power law model was necessary even in presence of an extra absorption.
The possibility of a presence of circumnuclear ionized gas was suggested by 
\cite{Sambruna98} who reported evidence of an absorption feature around 0.6 keV with a neutral hydrogen 
column density $N_{\rm H}\simeq 10^{21}$ cm$^{-2}$, however this feature was not confirmed
by a subsequent spectroscopic {\it XMM-Newton} observation (\cite{Blustin04}). 
Finally, the {\it Chandra} detection of a diffuse soft X-ray radiation around \pks, interpreted as a 
thermal emission of the host galaxy on kpc scale, was reported by \cite{Donato03}.

This interesting source was observed on several occasions by the \swift satellite 
(\cite{Gehrels04}) from April to June 2005. 
In this paper we present the results of a detailed spectral analysis of the X-Ray 
Telescope (XRT, \cite{Burrows05}) and Ultraviolet/Optical Telescope (UVOT, \cite{Roming05}) 
data, confirming that the X-ray spectrum of this HBL object shows a well established 
curvature similar to that found in other sources of the same type. 
In Section \ref{obs} the observations and the data reduction 
are presented, in Section \ref{xrt} we describe the \swift XRT spectral analysis and 
Section \ref{sax} is dedicated to the {\it Beppo}SAX data analysis. Finally the results are discussed 
in Section \ref{discussion}. Throughout this paper errors are quoted at the 90\% confidence level for one  
parameter of interest ($\Delta\chi^{2}=2.7$) unless otherwise specified. 

%-------------------------------------------------------------
   \begin{figure}
   \centering
    \includegraphics[width=8.cm,angle=-90]{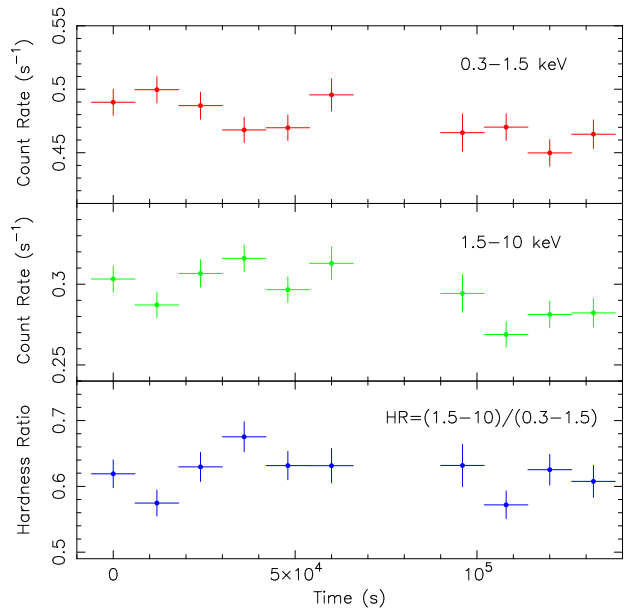}
     \caption{\swift XRT 0.3--1.5 keV (upper panel) and 1.5--10 keV (middle panel) light curves of \pks. 
              In the lower panel the corresponding Hardness Ratio (HR) is plotted. Data are binned to 
              12 ks resolution and error bars indicate statistical uncertainties at the 1$\sigma$ level.
             }
         \label{hardness}
   \end{figure}
%-------------------------------------------------------------

\section{Observations}
\label{obs}

As part of a \swift key project dedicated to the observation and monitoring of a 
sample of Blazars \pks~ was pointed eleven times over the period  April-June 2005. 
The journal of these observations is given in Table \ref{tab1} where we also report the 
net exposures with the XRT and UVOT instruments. On May 22 a deep (about 40 ks) 
XRT exposure of \pks~ was taken.
The exposure times in the Burst Alert Telescope (BAT, \cite{Barthelmy05}) 
were  not sufficient to detect a source with a typical intensity lower than 2 mCrab like \pks.

\begin{table*}[t]
\caption{Best fit spectral parameters of the log-parabolic model for the XRT 
(first section) and \sax (second section) observations of \pks. 
Numbers in parenthesis are statistical errors at the 90\% confidence level.
Fluxes in the 2--10 keV energy band are corrected for Galactic absorption.
}
\label{tab2}
\begin{center}
\begin{tabular}{llcccccc}
\hline
\hline
Instrument& Date     & $K$  & $a$ & $b$ & $E_p$ & $F_{2-10}$ & $\chi^2_r$/dof  \\
   &  &  & &  & (keV) & (erg cm$^{-2}$ s$^{-1}$) &  \\
\hline
\swift XRT & 2005 April 26 & 1.41 (0.05)~10$^{-2}$ & 1.81 (0.07) & 0.18 (0.13) & 3.4  &
4.0$\times$10$^{-11}$ & 0.75/118 \\
\swift XRT & 2005 May 21 & 1.48 (0.04)~10$^{-2}$ & 1.71 (0.05) & 0.38 (0.10) & 2.4  & 
4.0$\times$10$^{-11}$ & 0.90/196 \\
   &         & 1.46 (0.04)~10$^{-2}$ & 1.78 (0.03) & 0.18 frozen   & 3.5  & 
4.2$\times$10$^{-11}$ & 0.94/197 \\
\swift XRT & 2005 May 22 & 1.66 (0.02)~10$^{-2}$ & 1.90 (0.02) & 0.18 (0.04) & 1.9  & 
4.1$\times$10$^{-11}$ & 1.03/408 \\
\swift XRT & 2005 May 29 & 1.31 (0.04)~10$^{-2}$ & 1.77 (0.06) & 0.34 (0.11) & 2.2  & 
3.3$\times$10$^{-11}$ & 1.01/181 \\
    &        & 1.29 (0.04)~10$^{-2}$ & 1.83 (0.04) & 0.18 frozen   & 2.7  & 
3.5$\times$10$^{-11}$ & 1.03/182 \\
\swift XRT & 2005 June 24 & 1.52 (0.05)~10$^{-2}$ & 1.79 (0.06) & 0.26 (0.11) & 2.5  & 
4.1$\times$10$^{-11}$ & 1.09/183 \\
\hline
{\it Beppo}SAX LECS+MECS & 1999 Febr. 20 & 0.77 (0.07)~10$^{-2}$ & 1.57 (0.10) & 0.48 (0.10) 
&  2.8  & 2.3$\times$10$^{-11}$ & 1.23/55  \\
{\it Beppo}SAX LECS+MECS & 1999 April 07 & 0.69 (0.06)~10$^{-2}$ & 1.82 (0.10) & 0.42 (0.10) 
& 1.6  & 1.5$\times$10$^{-11}$ & 0.91/55 \\
\hline

\multicolumn{6}{c} { }
\end{tabular}
\end{center}
\end{table*}

\subsection{XRT data reduction}

To obtain a signal with sufficient statistics for a detailed spectral analysis 
we only considered observations longer than 5 ks. In particular, the X-ray spectrum 
accumulated during the longest XRT observation of May 22 has an 
excellent photon statistics and allowed us to perform a very accurate spectral analysis.
All XRT observations were carried out using the most sensitive Photon Counting readout mode 
(see \cite{Hill} for a description of readout modes).
The XRT data were processed with the XRTDAS software package (v.1.8.0). 
Event files were calibrated and cleaned with standard filtering criteria with the {\it xrtpipeline} 
task using the latest calibration 
files available in the \swift CALDB distributed by HEASARC. Events in the energy range 
0.3--10 keV with grades 0--12 were used in the analysis (see \cite{Burrows05} for a definition of 
XRT event grades).

The source count rate was high enough to cause some photon pile-up in the inner 6 pixel 
($\sim 14''$) radius circle within the peak of the telescope Point Spread Function (PSF), as 
derived from the comparison of the observed PSF profile with the analytical model reported by 
\cite{Moretti05}. We thus avoided pile-up effects selecting events within an annular region 
with an inner radius of 6 pixels and an outer radius of 30 pixels.
The background was extracted from a nearby source-free circular region of 50 pixel radius. 
Ancillary response files for the spectral analysis were generated with the {\it xrtmkarf} task 
applying corrections for the PSF losses and CCD defects. The latest response matrices (v.~008) 
available in the \swift CALDB were used.
The spectrum was binned to ensure a minimum of 20 counts per bin, and energy channels between 
0.4 keV and 0.6 keV were excluded to avoid undesired effects on the measured spectral parameters due 
to residual instrumental features (Campana et al.~2006).

\subsection{UVOT data reduction}
\label{uvot}

UVOT observations with good exposure times were available for a number 
of pointings (see Table \ref{tab1}).
Sky corrected images were derived from the \swift archive and aperture 
photometry was made with UVOTSOURCE using a 6$''$ (12 pixels) radius
for the $V, B, U$ filters and 12$''$ for the $W1, M2$ and $W2$ filters.
Count rates in all the filters were always less than 5 counts/s, well below 
the pile-up threshold (10$-$15 counts/s).

The source showed no appreciable variations in all the pointings, with 
an average value $<B>=16.9$. We report here only the data regarding 
the May 21 observation, which is the longest and is nearly simultaneous 
with the longest XRT observation of May 22.

The host galaxy of this source is a giant elliptical that has been 
extensively studied (e.g. by Falomo et al. 1995). 
To estimate the galaxy contribution within our 6$''$ aperture
we integrated the best fit De Vaucoulers profile published by them
(their Fig.4) obtaining $R$=15.7. Assuming the typical colors for elliptical
galaxy ($B-V$=0.96, $V-R$=0.61; Fukugita et al. 1995) we derive
$V$=16.3 and $B$=17.2. The AGN luminosity is therefore just 25\% of the observed
flux in these bands. For this reason we considered only photometric data in the U and UV bands 
where the host galaxy contribution is smaller.
The measured magnitudes in the U and UV filters were $U=16.5$, $UVW1=16.4$, $UVM2=16.7$, $UVW2=16.6$.
As the zero point uncertainty in the optical bands of 
UVOT is about 0.1 mag (about 10\% in flux) an accurate estimate of the spectral 
shape is not currently feasible.

\section{The X-ray spectrum}
\label{xrt}

To study the X-ray spectral distributions of \pks~ we first considered the deep 40 ks exposure
performed on May 22. In Fig.~\ref{hardness} we show the 0.3--1.5 keV (upper panel) and 
1.5--10 keV (middle panel) light curves of \pks, together with the corresponding hardness ratio 
(lower panel). From the figure is apparent that no significant temporal and spectral variability 
was present during the observation. 
We thus performed the spectral analysis using the events accumulated in the entire 
duration of the observation. We adopted the following two spectral models, a 
single power law:
\begin{equation}
F(E) = K ~E^{-\alpha} ~~~~{\rm photons~ cm}^{-2} {\rm s}^{-1} {\rm keV}^{-1}
\end{equation}
and a log-parabolic law (\cite{Mas04a}):
\begin{equation}
F(E) = K ~E^{-(a + b Log E)} ~~~~{\rm photons~ cm}^{-2} {\rm s}^{-1} {\rm keV}^{-1}
\label{eq.logpar}
\end{equation}
where the parameter $a$ is the photon index at 1 keV while $b$ measures the curvature of the parabola. 
The latter model has the property of describing curved spectra with only one additional 
parameter with respect to the single power law. The log-parabolic model is also very useful for 
the estimation of the energy and the flux of the SED peak, simply given by, respectively:
\begin{equation}
E_p = 10^{(2-a)/2b} ~~~~{\rm  keV},
\label{eq.logpar2}
\end{equation}
\begin{equation}
\nu_p F(\nu_p) = (1.60\times10^{-9})~K~10^{(2-a)^2/4b} ~~~{\rm erg~ s^{-1}~ cm^{-2}}.
\label{eq.logpar3}
\end{equation}
The log-parabolic law generally fits well the X-ray spectra of HBL sources, as 
shown by Massaro et al.~(2004a,b) for the cases of the X-ray spectra of Mkn~421 
and Mkn~501.
The relations between the parameters of synchrotron and inverse Compton 
radiation emitted by an electron population with a similar energy distribution
have been recently investigated by Massaro et al.~(2006). 

\begin{figure}
   \centering
   \includegraphics[width=5.6cm,angle=-90.]{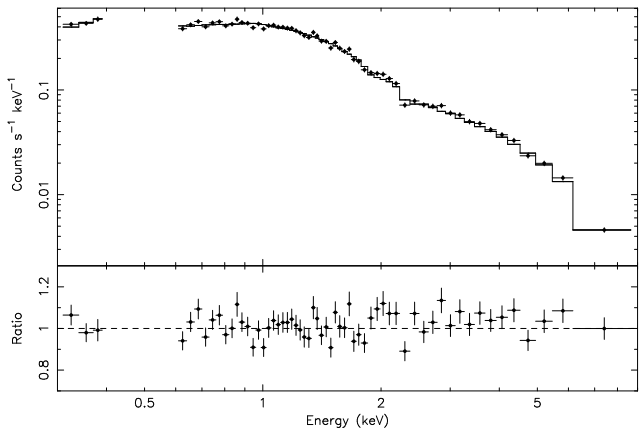}
      \caption{Best fist spectrum and residuals for the log-parabolic
               model of 0.3--10 keV XRT data (2005, May 22). Data points 
               between 0.4 and 0.6 keV were removed to avoid residual 
               calibration uncertainties in this energy range.
               }
         \label{fitlogpar}
   \end{figure}

When considering curved spectra a parameter that needs special attention is the amount 
of low energy neutral hydrogen absorbing column density $N_{\rm H}$. 
We first fitted the background subtracted 0.3-10 keV spectrum with a single power law model with the 
absorption column density fixed at the known Galactic value of 
$N_{\rm H}$=2.49$\times$10$^{20}$ cm$^{-2}$ (\cite{Murphy96}). We obtained a photon index of 
1.96$\pm$0.02 with reduced $\chi^2_r$ of 1.15 with 409 degree of freedom (dof).
From the inspection of the residuals a clear systematic effect reflecting the presence of 
spectral curvature was observed. We thus adopted the log-parabolic model with the same amount of $N_{\rm H}$.
The model provided a very good fit with $\chi^2_r$=1.03 (408 dof), and a $\chi^2$ decrease of 50 for 
only one additional parameter. 
The $F$ test gives a probability of about 10$^{-11}$ that this improvement is due to chance. 
Figure \ref{fitlogpar} shows the best fit spectrum with this model and the residuals.

%-------------------------------------------------------------
%-------------------------------------------------------------

The same log-parabolic spectral law was then also applied to the analysis of the observations 
of shorter duration. The resulting best fit parameters, the SED peak energies $E_p$ and fluxes in the 
2--10 keV band are given in Table \ref{tab2}. 
We note that for these shorter observations the parameters $a$ and $b$ are characterized by
statistical uncertainties much larger than those of the May 22 observation.  
In two cases (April 26 and June 24) the value of $b$ was found to be consistent with the one measured on 
the May 22 deep observation, while on May 21 and 29 it was larger ($b\sim 0.3-0.4$).
For these two latter pointings, best fits with the $b$ value frozen at 0.18 (the value May 22 value) 
gave very small increases of the $\chi^2$ values (see Table \ref{tab2}), indicating that the larger 
curvature values measured are not statistically significant.

We also verified the possibility to have an additional intrinsic absorption 
either in the nuclear environment of \pks~ or in the host elliptical galaxy. We recall 
that the exclusion of data points between 0.4 and 0.6 keV do not 
allow an accurate estimation of the absorption. 
A simple power law best fit with a free $N_{\rm H}$ for the May 22 observation 
resulted in a slightly larger absorption column density of (4.2$\pm$ 0.6)$\times$10$^{20}$ cm$^{-2}$. 
However, the $\chi^2$ was found comparable to the fit with $N_{\rm H}$ fixed at the Galactic value 
($\chi^2_r$/dof=1.14/408) and, again, the residuals showed a clear spectral curvature. 
The May 22 spectrum was also fit with the log-parabolic model leaving $N_{\rm H}$ free to vary. The fit 
was good ($\chi^2_r$/dof=1.03/407) and statistically equivalent to the case with fixed Galactic absorption. 
We obtained an absorption column density of 
(2.1$\pm$ 1.0)$\times$10$^{20}$ cm$^{-2}$, in agreement 
with the Galactic value, and a curvature $b=0.21\pm 0.10$.

Therefore, we conclude that the X-ray spectrum of \pks~ is intrinsically curved and that 
it is well described by a log-parabolic model with no excess absorption 
required, like the two well known HBL sources Mkn~421 and Mkn~501.

\section{\sax observations}
\label{sax}

In this section we present a re-analysis of \sax data considering the log-parabolic model, 
not used by \cite{Cost01a}, to evaluate the spectral curvature and compare it to that seen by the XRT 
in different brightness states.

%-------------------------------------------------------------
   \begin{figure*}[t]
   \centering
   \includegraphics[width=9.5cm,angle=-90]{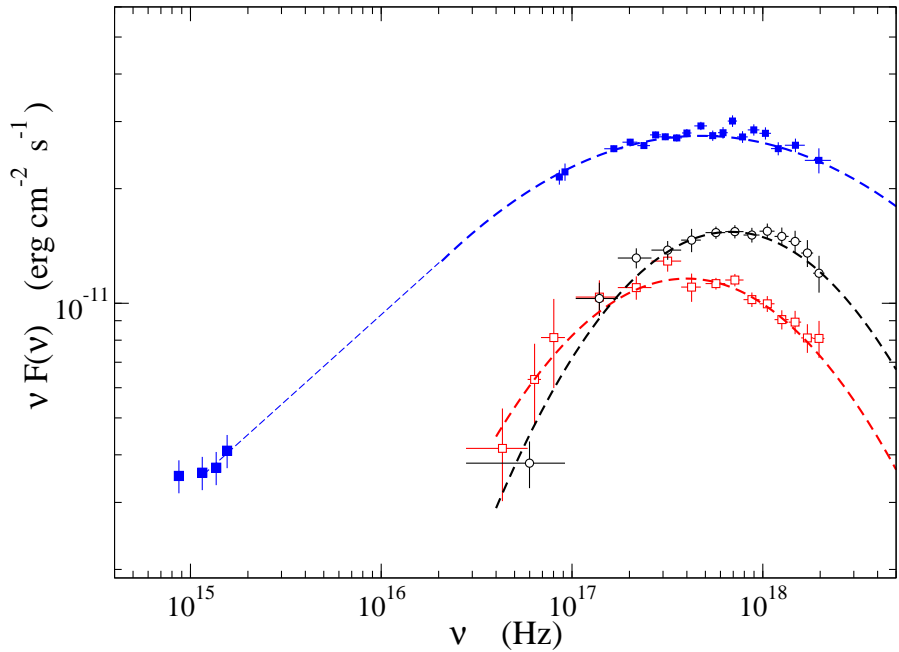}
      \caption{Optical, UV and X-ray Spectral Energy Distribution of \pks~ from \sax (1999 February 20, 
               open circles; 1999 April 07: open squares), \swift XRT 
               (2005 May 22, filled circles) and \swift UVOT (2005 May 21: filled squares) observations. The 
               thick dashed lines represent the log-parabolic best fit models to the X-ray data points. The 
               thin dashed line is a power law extrapolation of the UV spectral data to the soft X-ray band.
               Error bars indicate statistical uncertainties at the 1$\sigma$ level. 
               The host galaxy contribution has not been subtracted from UVOT data (see Sect.~\ref{uvot}).
              }
         \label{sed}
   \end{figure*}
%-------------------------------------------------------------

In the two \sax observations in 1999 (February 20 and on April 7) \pks~ was too faint to be detected by the 
PDS instrument and therefore we analyzed only LECS and MECS data. 
Events for spectral analysis were selected in circular regions of 6$'$ and 4$'$ and in 
the energy bands 0.1--2.0 keV and 2.0--10.0 keV for the LECS and MECS, respectively. 
Background spectra were taken from the blank field archive at the ASI Science Data Center.
In Table \ref{tab2} we reported also the log-parabolic best fit parameters of these \sax observations. 

We first note that in 1999 the typical 2--10 keV flux of \pks~ was about a factor of 2 lower
than in 2005. This change of luminosity was accompanied by a different spectral curvature ($b\simeq0.45$) 
while the peak energy appears more stable. To verify that this change of curvature 
is real we fit the {\it Beppo}SAX data with $b$ kept frozen to the value measured by \swift in the May 22 
observation.  
We found an increase of $\chi^2$ corresponding to an $F$-test probability of 
$4\times10^{-6}$ and $2\times10^{-5}$ for the observation of February 20 and April 7, respectively. 
These results make us confident that the change of curvature is very significant.
We also verified the possibility to have an extra-absorption 
in the local frame of \pks. 
The two \sax 1999 spectra were fit with the log-parabolic model with free $N_{\rm H}$. In both pointings we 
found an absorption column density of $N_{\rm H}=(3.2\pm 1.2)\times10^{20}$ cm$^{-2}$, consistent with the 
Galactic value, and curvatures $b=0.39\pm0.18$ and $b=0.32\pm0.20$ for the February 20 and April 7 observations, 
respectively. Moreover, the fits were statistically equivalent with the ones with fixed absorption. 
We conclude that the X-ray spectrum of \pks~ was intrinsically curved also in the two \sax 1999 
observations and is well described by a log-parabolic model with Galactic absorption.

\section{Discussion}
\label{discussion}

Our spectral analysis of the recent series of \swift XRT observations of \pks~ has shown 
that the X-ray spectrum of this BL Lac object is characterized by a significant curvature that is well 
fitted by a log-parabolic law with a curvature parameter and peak energy similar to those 
of other HBL sources (e.g. \cite{Mas04a, Gio02c,Tramacere06}). 
Blustin et al.~(2004) summarized some historical X-ray data of \pks~ since its first X-ray 
observation of March 1979 and found that the 2--10 keV flux in this source 
was always within the range ($\sim$1.5 -- $\sim$4.5)$\times$10$^{-11}$ erg cm$^{-2}$ s$^{-1}$.
The \swift XRT observations in the spring 2005 described in this paper confirm this rather 
stable behavior with observed fluxes close to the historical maximum. 
This relative stability at X-ray frequencies of \pks~ contrasts with the behavior of the Mkn~421 
and Mkn~501 (\cite{Fossati00}; Massaro et al.~2004a,b) whose X-ray fluxes varied well over an order of 
magnitude.
Of course, the sampling of this historic X-ray data set is rather poor and we cannot exclude 
the occurrence of large outbursts on a time scale of a few months, possibly associated with 
spectral changes.

Figure \ref{sed} shows the SEDs of \pks~ in the UV to X-ray range for the long May 22 pointing and the
two \sax observations. 
The main differences between the recent high state with respect to the previous data is the 
well apparent decrease of the curvature whereas the peak energy is around the same values.
Moreover, the UV data can be connected to the low energy X-ray points by a power law interpolation 
(thin dashed line) suggesting that they come from the same component.
Simultaneous UV data are not available for the \sax observations: however, a low-energy
extrapolation of the X-ray SEDs indicates that at those epochs the UV emission of \pks~ should 
have been lower than that found in 2005.
A recent extensive study of a large set of X-ray observations of Mkn~421 (Massaro et al. 2004a, 
Tramacere et al. 2006b) has shown that this HBL source shows a positive correlation between the  
energy and the flux at the SED peak.
In the case of \pks~ we do not observe a significant increase of the peak energy with flux. 
This finding can be an indication that statistical acceleration works with different 
efficiencies in these two sources.

\pks~ is a good candidate for a detection in the TeV range, as already noticed by 
\cite{Costam02}.
In a Synchrotron Self-Compton scenario a brightening at X-ray frequencies 
often corresponds to an enhanced TeV luminosity. 
It is therefore very useful to organize a monitoring program of the X-ray 
flux in the next months to trigger a possible TeV detection. 
The study of the evolution of both synchrotron and inverse Compton components 
in the SED will be very useful to understand the physical conditions 
in the nuclear region and, when compared with other TeV BL Lacs, to derive 
a more general picture of this class of sources.
Thanks to the wide field of view of the BAT instrument, and to its fast 
pointing capability with the UVOT and XRT narrow field telescopes, 
\swift is the best suited satellite to perform such a program.

\begin{acknowledgements}

We are grateful to the referee for his/her useful comments and suggestions.
The authors acknowledge the financial support for
ASDC by the Italian Space Agency (ASI) and for the Phys.
Dept. by Universit\`a di Roma La Sapienza. 

\end{acknowledgements}

\end{document}